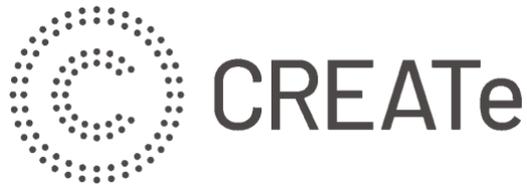

# Accuracy of training data and model outputs in Generative AI: CREATe Response to the Information Commissioner's Office (ICO) Consultation

Zihao Li[*], Weiwei Yi[†], Jiahong Chen[‡]

The accuracy of Generative AI is increasingly critical as Large Language Models (LLMs) become more widely adopted. Due to potential flaws in training data and hallucination in outputs, inaccuracy can significantly impact individuals' interests by distorting perceptions and leading to decisions based on flawed information. Therefore, ensuring these models' accuracy is not only a technical necessity but also a regulatory imperative. ICO's call for evidence on the accuracy of Generative AI marks a timely effort in ensuring responsible Generative AI development and use.

CREATe, as the Centre for Regulation of the Creative Economy based at the University of Glasgow, has conducted relevant research involving intellectual property, competition, information and technology law. We welcome the ICO's call for evidence on the accuracy of Generative AI, and we are happy to highlight aspects of data protection law and AI regulation that we believe should receive attention. More information about CREATe can be found at https://www.create.ac.uk/.

1. **Do you agree with the analysis presented in this document?**

The analysis presented a comprehensive understanding of data accuracy in Generative AI (GenAI). However, there are five overarching issues that have been overlooked or underestimated, notably: (1) merely relying on the disclosure of statistical accuracy of the

---

GenAI model is insufficient, since it could lead to an "**Accuracy Paradox**";[1] (2) increasing the accuracy of inputs, models, and outputs often comes with the cost of privacy, especially in GenAI context; (3) overreliance on developers' and deployers' accuracy legal compliance is not pragmatic and is overoptimistic, which could ultimately become a burden for users with the tendency of using dark pattern;[2] (4) it is very difficult to anticipate the specific application scenarios of Gen AI, especially when it comes to Large Language Models (e.g., ChatGPT, Co-pilot and Gemini); (5) the accuracy of training data cannot directly translate to the accuracy of output, which may pose significant risks to individuals. This is particularly evident in cases of hallucination. The following part will articulate these five issues:

(1) **Accuracy Paradox**: It is agreed that disclosure of statistical accuracy of the generative AI model is necessary. However, if it is the only measure that GenAI developers are required to undertake, it could be counterproductive due to the Accuracy Paradox. The result of disclosing statistical accuracy is likely to lead to a new accuracy race among providers. Users will highly rely on this because it is the only metrics that developers need to unveil. However, merely disclosing the accuracy rate and improving the accuracy of the models through new data and algorithms are insufficient, because the more accurate the model is, the more users will rely on it, and thus be tempted not to verify the answers, leading to greater risk when stochastic parrots and hallucinations appear.[3] Even if the model is highly accurate, it cannot provide a 100% trustworthy answer because LLMs only predict the likelihood of the words occurring, without comprehending the materials they are working with. If 0.1% or 0.2% of the answers in a specific field are untrue, this will pose an enormous difficulty for users in identifying the authenticity of the answers. The risk is beyond measure if users encounter these problems in particularly sensitive areas such as healthcare or the legal field. Even if utilizing real-time internet sources, the trustworthiness of LLMs may remain compromised, as exemplified by factual errors in the new Bing's launch demo.[4]

---

[1] Accuracy paradox refers to the unintended consequences of solely relying on the disclosure of a model's statistical accuracy, which can lead to a misleading sense of reliability among users. As accuracy metrics improve, users may overly trust the AI outputs without sufficient verification, increasing the risk of accepting erroneous information. This paradox describes how higher reported accuracy can paradoxically lead to greater risks associated with overreliance on these metrics.
[2] Dark pattern refers to design choices that are meant to manipulate users into doing things they might not otherwise do. In this context, GenAI developers and deployers could use such manipulative design to shift the responsibility for data accuracy onto users. They could exploit the overreliance on their compliance with accuracy standards to subtly coerce users into continuously verifying and updating their personal data to maintain the AI's accuracy.
[3] Zihao Li, 'Why the European AI Act Transparency Obligation Is Insufficient' (2023) Nature Machine Intelligence, https://doi.org/10.1038/s42256-023-00672-y

(2) **Implications of accuracy-privacy trade-off for GenAI contexts**: Increasing accuracy of the inputs, model, and outputs often comes at the cost of privacy.[5] This involves not only technical identifiability of the individuals involved[6], but also societal risks such as more accurate and precise targeting for commercial purposes, social sorting, and group privacy implications. While the ICO's analysis rightly connects the accuracy principle with the AI system's purpose, asserting that the appropriate level of privacy depends on the system's intended use, it should more explicitly require developers to demonstrate that improving accuracy does not compromise other interests. The risk of not specifically spelling that out is that some developers might try to justify their over-collection and over-analysis of data in the name of improving accuracy.

(3) **Overreliance on developers' and deployers' accuracy compliance would turn into users' burden**: Building on the previous points, GenAI developers might shift such responsibility for submitting accurate personal data to users. By playing up the risks relating to inaccuracy on the data subjects, the GenAI developers could force and nudge data subjects to provide accurate and up to date personal data, sacrificing their de facto privacy to benefit the AI developers' business interests. Therefore, in the context of GenAI, it is crucial to distinguish between scenarios in which the AI generates direct personal information of a specific person based on the provided training data, and in which it only displays statistical data of a group. In the first scenario, ICO might want to offer guidelines regarding the design of the presence of information and prohibit coercive framing during data collection that creates unjustifiable burdens on and forces data subjects to periodically verify data accuracy. In the second scenario, such purposes and forms of disclosure should be clearly communicated to data subjects to facilitate informed decision-making of data sharing and rectification. However, in both cases, it may be worthwhile for the ICO to consider whether it could recommend or even mandate that GenAI systems provide sources of references (e.g., via hyperlinks) that are most influential, or when the references are a few in number. This approach would allow end-users of the generative AI to verify accuracy themselves, potentially mitigating the risks of user overreliance and reducing the democratic risks associated with generative AI dictating the narrative of facts.[7]

---

[4] Kif Leswing, 'Microsoft's Bing A.I. Made Several Factual Errors in Last Week's Launch Demo' *CNBC* (14 February 2023) <https://www.cnbc.com/2023/02/14/microsoft-bing-ai-made-several-errors-in-launch-demo-last-week-.html> accessed 10 May 2024.
[5] Jiahong Chen, 'The Dangers of Accuracy: Exploring the Other Side of the Data Quality Principle' (2018) 4 European Data Protection Law Review 36.
[6] Benjamin Zi Hao Zhao, Mohamed Ali Kaafar and Nicolas Kourtellis, 'Not One but Many Tradeoffs: Privacy Vs. Utility in Differentially Private Machine Learning', *Proceedings of the 2020 ACM SIGSAC Conference on Cloud Computing Security Workshop* (ACM 2020) <https://dl.acm.org/doi/10.1145/3411495.3421352> accessed 8 May 2024.

(4) **Content moderation as a tool to mitigate inaccuracy and untrustworthiness**: While it is recognised that the connection of the accuracy principle with the system's purpose is a right direction, it is difficult anticipate the specific application scenarios of GenAI, especially when it comes to LLMs. Due to the rapid development of LLMs and their expanding capabilities, the multifaceted nature of LLMs introduces complexities in linking their design purpose to its application context,[8] especially when downstream developers are capable of building systems with a different purpose. Such usage of GenAI content is often out of developers and even deployers' control. Therefore, when it comes to the accuracy in such general-purpose model, the additional obligation in accuracy should focus on the content and output moderations. As a critical role in ensuring the accuracy, reliability, and trustworthiness of GenAI, content moderation could filter flawed or harmful content, which involves refining detection methods to distinguish and exclude incorrect or misleading information from training data and model outputs. Moreover, it could reduce potential bias, which entails implementing strategies to identify and mitigate prejudiced data patterns, ensuring that AI outputs are fair and representative of diverse perspectives.

(5) **Accuracy of training data cannot directly translate to the accuracy of output, especially in the context of hallucination**: As the analysis recognised, the key reason for inaccuracy in output of GenAI is hallucinations. This issue largely arises because the text production method of LLMs is to reuse, reshape, and recombine the training data in new ways and patterns to answer new questions, often ignoring the problem of authenticity and trustworthiness of the answers. Although the majority of answers are high-quality and true, the content of the answers remains subject to further verification. The likelihood of inaccuracy is probabilistic and inevitable. Even though most training data is reliable and trustworthy, the essential issue remains that the recombination of trustworthy data into new answers in a new context may lead to untrustworthiness, as the trustworthiness of information depends on the context and circumstances. When taken out of context, the trustworthiness of the outputs may no longer be maintained. Therefore, merely focusing on the accuracy of input personal data is insufficient, since it cannot transform to accuracy of output that could have significant impact on individuals. We suggest that the ICO also pay more attention to authentication, verification and content moderation mechanisms in the context of GenAI. Both could effectively address the issue of output inaccuracy.

---

[7] Li (n 3).
[8] Magali Eben and others, 'Priorities for Generative AI Regulation in the UK: CREATe Response to the Digital Regulation Cooperation Forum (DRCF)' <https://zenodo.org/record/8319662> accessed 9 May 2024.

## Q6: What technical and organisational measures can organisations use to improve the statistical accuracy of generative AI models?

There are several measures that organisations can use to improve statistical accuracy or minimise inaccuracy. Firstly, zero-shot Fact Verification can effectively enhance the accuracy of generative AI models by automatically generating diverse claims from existing evidence, enabling the models to verify output for a broad range of scenarios without human annotation.[9] Furthermore, it is observed that implementing automated error correction systems within LLMs, such as the Chinese Large Language Model Kimi, significantly enhances model accuracy, precision, and recall.[10] Such systems actively detect and correct errors in real-time, adapting to evolving language patterns without extensive retraining. The integrated feedback loops allow the model to adapt based on user interactions, continuously refining error detection and correction processes, thereby reducing the overall error rate and enhancing system responsiveness and adaptability.[11] Furthermore, dynamic real-time information injection can, to some extent, address issues of content hallucination and data relevancy, thereby improving the accuracy of model outputs.[12] This approach involves dynamically collecting and integrating current data from credible sources into model prompts, ensuring that the generated content is both current and factually correct.[13]

**Reference**
Chen J, 'The Dangers of Accuracy: Exploring the Other Side of the Data Quality Principle' (2018) 4 European Data Protection Law Review 36

---

[9] Liangming Pan and others, 'Zero-Shot Fact Verification by Claim Generation' (arXiv, 30 May 2021) <http://arxiv.org/abs/2105.14682> accessed 9 May 2024.
[10] Wai-lam Cheung and Chiu-Ying Luk, 'Implementing Automated Error Correction and Feedback Loops in Kimi, A Chinese Large Language Model' (24 April 2024) <https://osf.io/7vpxr> accessed 9 May 2024.
[11] Qian Ouyang, Shiyu Wang and Bing Wang, 'Enhancing Accuracy in Large Language Models Through Dynamic Real-Time Information Injection' (26 December 2023) <https://www.preprints.org/manuscript/202312.1987/v1> accessed 9 May 2024.
[12] ibid.
[13] ibid.